\documentclass[prl,twocolumn,showpacs,groupedaddress]{revtex4}

\usepackage{amssymb}
\usepackage{amsmath}
\usepackage{dcolumn}
\usepackage{bm}
\usepackage{graphicx}

\newcommand{\ket}[1]{|#1\rangle}

\begin{document}

\title{Mapping electron delocalization by charge transport spectroscopy in an artificial molecule}

\author{M. R. Gr{\"a}ber$^{\S}$}
\author{M. Weiss}
\author{D. Keller}
\author{S. Oberholzer}
\author{C. Sch{\"o}nenberger}

\email{christian.schoenenberger@unibas.ch} \affiliation{Department of Physics,
University of Basel, Klingelbergstr.~82, CH-4056 Basel, Switzerland }

\date{\today}

% ------------------------------------------------------------------
% Abstract
% ------------------------------------------------------------------

\begin{abstract}
In this letter we present an experimental realization of the
quantum mechanics textbook example of two interacting electronic
quantum states that hybridize forming a molecular state.
In our particular realization, the quantum states themselves
are fabricated as quantum dots in a molecule, a carbon nano\-tube.
For sufficient quantum-mechanical interaction (tunnel coupling) between
the two quantum states, the molecular wave\-function is a superposition of the two
isolated (dot) wave\-functions. As a result, the electron becomes
delocalized and a covalent bond forms.
In this work, we show that electrical transport can be used as a sensitive probe
to measure the relative weight of the two components in the superposition state
as a function of the gate-voltages.
For the field of carbon nano\-tube double quantum dots, the findings represent
an additional step towards the engineering of quantum states.

\end{abstract}

% Use showkeys class option if keyword display desired

\maketitle

% ------------------------------------------------------------------
% Main text
% ------------------------------------------------------------------

% ------------------------------------------------------------------
%\section{ Introduction}
% ------------------------------------------------------------------

In the quantum world particles such as electrons behave as extended
objects with the character of a wave. The combination of wave-like
behaviour on the one hand and interactions on the other hand leads
to `localized' electron waves, also called quantum states. When two
quantum states overlap, quantum interference results in a
superposition state with new qualities. In particular, a molecular
bond may emerge from the constructive interference of the quantum
states. Engineered double quantum dots provide an experimental
platform enabling one to control this prototype of molecular
formation
\cite{Livermore,Blick,Hatano,Fasth,dicarlo,Holleitner,pioro}. In
this letter, the molecular states are realized in a double quantum
dot device gate-defined in a single-walled carbon
nano\-tube (SWCNT)~\cite{Mason,sapmaz,graebermol}. We show that, by analyzing
the electrical conductance through the device, it is actually
possible to map the electron delocalization of the covalent chemical
bond formed between the two quantum dots.

\begin{figure}[h]
\includegraphics*[width=\linewidth]{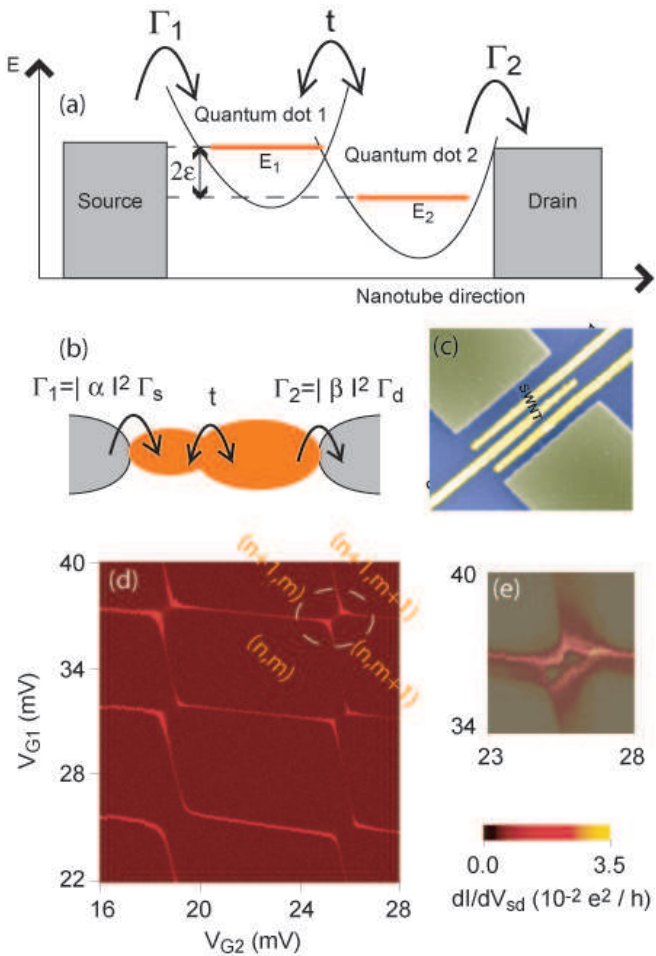}
\caption{\label{figure1} (a)~Schematic of the double
quantum dot toy model. Each dot contains a single level. For
sufficient tunnel coupling ($t\gtrsim \epsilon$) the two levels
hybridize. (b) Illustration of a molecular wave\-function
corresponding to the above scenario. The intrinsic charge transfer
rates to the leads $\Gamma_S, \Gamma_D$ are normalized by the
probabilities $\mid \alpha\mid^2, \mid \beta\mid^2$ of finding the electron on the
corresponding dot. (c)~Scanning-electron micrograph of a device. The
spacing between source and drain electrode amounts to 1.4~$\mu$m,
the gates are $\approx$ 150~nm wide. (d)~Colorscale plot of the
differential conductance $dI/dV_{sd}$ through the device at an
electron temperature of 50~mK. The high-conductance ridges define
the typical honeycomb-shaped charge stability map of a double
quantum dot. (e) Finite bias measurement of $dI/dV_{sd}$
($V_{sd}=500\;\mu$V) of the region marked by a dashed circle in
(d).}
\end{figure}

In Fig.~\ref{figure1}(a-c) we show the experimental realization of
our carbon nano\-tube double quantum dot device. A scanning electron
micrograph of such a device, showing source and drain contacts and three top-gate
electrodes in between can be seen in Fig.~\ref{figure1}(c). The
three top-gates can be used to adjust the electro\-static potential
landscape within the nano\-tube as depicted in Fig.~\ref{figure1}(a),
yielding the double quantum dot structure. Using the outer two gate
electrodes the difference of the level energies in the left and
right dot, the detuning $\epsilon = (E_{2}-E_{1})/2$ can be adjusted
as well. Transport takes place through a molecular state as
indicated in Fig.~\ref{figure1}(b), provided the de\-tuning is not
much larger than the tunnel coupling $t$, which in turn is proportional
to the overlap of the dot wave\-functions. One can say that the
de\-tuning $\epsilon$ determines the degree of localization of the
electron on the double quantum dot. Only for $\epsilon \approx 0$
the probability of finding the electron on the left or the right
dot will be comparable, for $\mid\epsilon\mid\gg 0$ the electron will
mainly be localized on one of the two dots.

In order to be more quantitative, we introduce a toy model of
molecular formation which is capable of describing all basic
features of our experiment. Our system consists of a left and a right quantum dot and
a single electron, which can be put into the dots. Left and right
dot are indexed by ´1´ and ´2´, respectively. We neglect spin and
consider only a single level per dot. The two levels of the two dots are
tunnel-coupled, characterized by the coupling energy $t$, which
we take as a real parameter. In the
basis of the single dot wave\-functions $\ket{\Phi_1}, \ket{\Phi_2}$
the system is described by the following Hamiltonian:

\begin{equation}
\mathbf{H}= \left(\begin{array}{cc}
E_1 & t \\
t & E_2 \\
\end{array} \right)\:\:.
\end{equation}

For the eigenvalues one obtains the energy of the bonding (+) and
the anti-bonding (-) orbital: $E^{\mp}=\Delta \pm \sqrt{\epsilon^2+
t^2}$, where we have defined $\Delta=(E_1+E_2)/2$ and the detuning
$\epsilon=(E_2-E_1)/2$, giving half the distance in energy between
the unperturbed energies $E_1$ and $E_2$. If there is no or only
little interaction ($t<<\epsilon$) between the two dots, the electron
will reside in either the left or the right dot in an
unperturbed eigen\-state $\ket{\phi_1}$ or $\ket{\phi_2}$,
respectively. For stronger coupling $t \agt \epsilon$, however, the
electron becomes delocalized. In general, the eigenstates
$\ket{\Psi}$ are then formed from a superposition
\begin{equation}
\ket{\Psi^{\pm}}=\alpha(\epsilon) \ket{\phi_1} \pm \beta(\epsilon)
\ket{\phi_2}\;\;\;,
\end{equation}
The probabilities of finding the delocalized electron on either the
left dot or the right dot are given by $\mid\alpha\mid^{2}$ and
$\mid\beta\mid^{2}$, respectively, with $\mid\alpha\mid^{2}+
\mid\beta\mid^{2}=1$. The coefficients $\alpha$ and $\beta$
depend on the detuning $\epsilon$; the probability of finding the
electron on the left dot being given by:
\begin{equation} \label{eq:coefficients}
\mid \alpha(\epsilon) \mid ^2
=\frac{1}{2}\,\left(1-\frac{\epsilon}{\sqrt{\epsilon^2+t^2}}\right) \:\:.
\end{equation}

Via the metallic source and drain electrodes, which are fabricated to the carbon nano\-tube
by nano-lithography (see Fig.~\ref{figure1}(c), one can
pass an electrical current through the device. This current will be
proportional to the joint probabilities of charge transfer between the
source and the quantum state in quantum dot $1$ (facing the source
electrode) and between the drain and the quantum state in dot $2$
(facing the drain electrode), exactly like in a single quantum dot.
The probabilities are commonly referred to as charge transfer rate $\Gamma_s$ and
$\Gamma_d$, for source and drain electrode, respectively. In the
case of a double quantum dot, when transport happens through a
molecular wave function, one has to weight this transfer rate with
the probability of the shared electron to be on the left or right
dot ($\mid\alpha\mid^2$ and $\mid\beta\mid^2$), respectively.
The transfer rates are then given by $\mid \alpha \mid ^2
\Gamma_{s}$ and $\mid\beta\mid^2 \Gamma_{d}$ and thus directly
reflect the molecular wave\-function. As was shown in
Ref.~\cite{graebermol}, the maximum linear differential
conductance for sequential tunneling through the molecular level can
be expressed as
\begin{equation} \label{eq:dIdV}
\frac{dI}{dV_{sd}}=\frac{e \Gamma(\epsilon)}{4 k_B T}\;\;,
\end{equation}
where
\begin{equation} \label{eq:Gamma}
\Gamma=\mid\alpha\beta\mid^{2}\frac{\Gamma_s \Gamma_d}{\alpha^2 \Gamma_s + \beta^2 \Gamma_d}\;\;.
\end{equation}
Equation~\ref{eq:dIdV} is valid if spin-degeneracy is lifted. Otherwise there is a slight
modification, which in linear transport only adds a correction factor
and does not affect the following analysis~\cite{Coish-Loss}.
Combining Eq.~\ref{eq:coefficients}-\ref{eq:Gamma}
yields $dI/dV$ as a function $\epsilon$. When comparing this dependence
with the experiment, the following three fitting parameters can in principle
be extracted: $t$ (actually $t^2$), $\Gamma_s$ and $\Gamma_d$. In practice, however,
an accurate coordinate transfer from gate-voltages to $\epsilon$ and $\Delta$
needs to be performed prior to this analysis, which also include additional fitting parameters in the
form of various capacitors~\cite{Coish-Loss}. Here, we illustrate in the following
that this procedure works and can yield sensible results.

\begin{figure}[t]
\includegraphics*[width=0.8\linewidth]{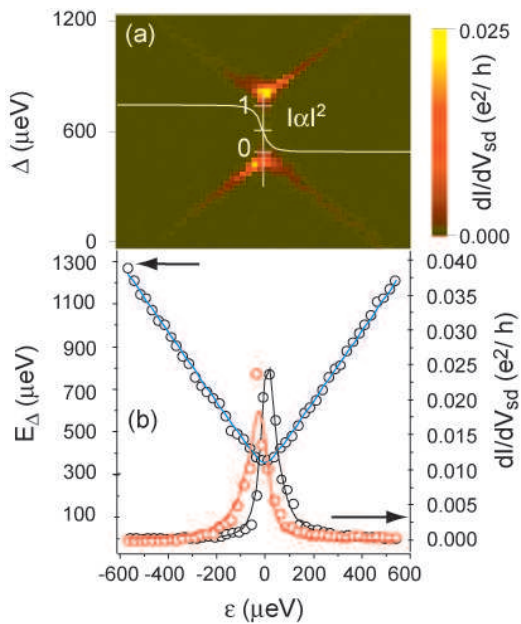}
\caption{\label{figure2} (a) Colorscale plot of the
differential conductance $dI/dV_{sd}$ through the device at 50~mK
versus $\epsilon$ and $\Delta$ in the triple point region marked by
the dashed circle in Fig.~\ref{figure1}(c). The inset shows the dependence of
$\mid\alpha\mid ^2$ on $\epsilon$ obtained from the spacing of the
wings (b) and from the magnitude of $dI/dV_{sd}$~(c). (b) Spacing
$E_{\Delta}$ of the two wings in (a) versus $\epsilon$, yielding
$t=30\,\mu$eV, and $dI/dV_{sd}$ versus $\epsilon$ for the
top (black) and the bottom (red) wing, yielding $t=34 and 40 \mu$eV,
respectively. Solid curves are fits to Eq.~\ref{eq:coefficients} and
Eq.~\ref{eq:dIdV}.}
\end{figure}

% ------------------------------------------------------------------
%\section{Experimental details}
% ------------------------------------------------------------------

Single-walled carbon nano\-tubes (SWNT) were grown by means of a
chemical vapor deposition (CVD) hydrogen-methane based process
\cite{1998-Kong,Comment-CVD} on a degenerately doped silicon
substrate, and located with a scanning-electron microscope (SEM).
The three 100~nm wide SiO$_2$/Ti/Pd top-gate electrodes and the
Pd/Al source and drain electrodes were defined by electron beam
lithography and physical vapor deposition \cite{graebermol}.
Electrical transport measurements using standard lock-in techniques
were performed in a He$^{3}$/He$^{4}$ dilution refrigerator at an
electronic temperature of $\approx$ 50~mK.

Figure~\ref{figure1}(d) shows the differential conductance through
the device at 50~mK in a colorscale representation. The bright
(high-conductive) lines define the charge-stability map of a
double quantum dot, often referred to as honeycomb
pattern~\cite{VanderWiel}. Within each cell the number (n,m) of
electrons residing on the two dots is constant. Increasing the
voltage applied to the left (right)~gate ($V_{G1(2)}$)
successively fills electrons into the left (right) dot, whereas
decreasing the gate voltages pushes electrons out of the dots. A
situation, where n~electrons are on the left~dot and m~electrons
on the right one, is denoted by (n,m). The curvature of the
high-conductive ridges indicates the formation of molecular
electronic eigenstates of the system, induced by a sufficiently
large tunnel coupling. In the following we will focus on the
ridges separating the (n,m), (n,m+1), (n+1,m), (n+1,m+1) charge
stability regions, marked by a dashed ellipse.

Next, the given gate voltages need to be transformed into the chemical
potential of either quantum dot. From Fig.~\ref{figure1}(d) it is
apparent that neither the vertical nor the horizontal boundaries
of the honeycomb cells run perfectly parallel to the gate voltage axis.
This is due to a mutual capacitance $C_m$ between the two dots, which
is small in carbon nano\-tube double dots as compared to other
planar GaAs heterojunction devices, but still not negligible for the
following analysis. Within an electrostatic model \cite{VanderWiel,Bruder},
the relations between gate voltages $V_{G1(2)}$ and chemical
potential $E_{1(2)}$ of the dots (including a constant offset) are
given by:
\begin{equation}
\left(\begin{array}{cc}
C_1 & -C_m \\
-C_m & C_2 \\
\end{array}\right)
\left(\begin{array}{cc}
\Delta E_1 \\
\Delta E_2 \\
\end{array}\right)=
\left(\begin{array}{cc}
C_{G1}\Delta V_{G1} \\
C_{G2}\Delta V_{G2} \\
\end{array}\right)\;\;,
\end{equation}  where $C_{1(2)}=C_{s(d)}+C_{G1(2)}+C_m$ denotes the total
capacitance of dot~1(2), with $C_{G1(2)}$ the capacitance of
top-gate~1(2) and $C_{s(d)}$ that of source and drain electrode,
respectively. From measurements of the double dot stability map at
finite bias (see Fig.~\ref{figure1}(e)), the dimensions of the
honeycomb cells, and the slope of the honeycomb boundaries, it is
possible to evaluate all capacitances of the device~\cite{graebermol,Coish-Loss}.
Since these capacitances may vary for large differences in gate
voltage, it is important to evaluate them for each triple point
region of interest individually. For the triple point region marked by a dashed
circle in Fig.~\ref{figure1}(d), we obtain $C_{G1} \approx 30$~aF,
$C_{G2} \approx 25$~aF, $C_1 \approx 51$~aF, $C_{2} \approx 52$~aF,
and $C_m \approx 8$~aF.

Figure~\ref{figure2}(a) shows a colorscale plot the differential
conductance of the above region versus $\epsilon$ and $\Delta$. The
bottom high-conductive wing corresponds to the situation when the
bonding state $\ket{E^{+}}$ is degenerate with the charge state
$\ket{n,m}$. Correspondingly, in the upper high-conductive wing,
$\ket{E^{+}}$ and $\ket{n+1,m+1}$ are degenerate. The spacing $E_{\Delta}$ of the two
wings in $\Delta$-direction is plotted versus $\epsilon$ in Fig.~\ref{figure2}(b). It is
described by $E_{\Delta}=U^{\prime}+2\sqrt{\epsilon^2+t^2})$~\cite{graebermol},
where $U^{\prime}$ denotes the electrostatic nearest-neigbour interaction.
Good fits to the data (thin solid curve) yield a tunnel coupling of
$t\approx 30\,\mu$eV and $U^{\prime}\approx 310\,\mu$eV.

\begin{figure}[t]
\includegraphics*[width=0.8\linewidth]{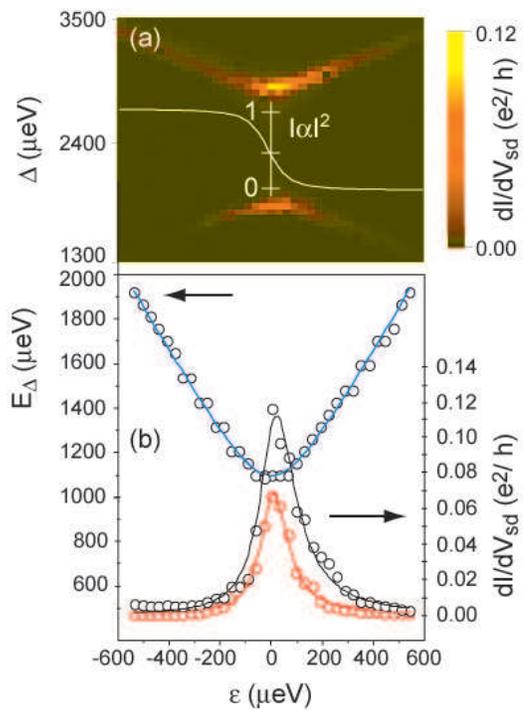}
\caption{\label{figure3} (a) Colorscale plot of the
differential conductance $dI/dV_{sd}$ in a strongly coupled triple
point region at 50~mK. The inset shows the dependence of $\mid\alpha\mid ^2$
on $\epsilon$ obtained from the spacing of the
wings (b) and from the magnitude of $dI/dV_{sd}$~(c). (b) Spacing
of the two wings and corresponding fit yielding $t\approx 110\,
\mu$eV, and Maximum conductance $dI/dV_{sd}$ with corresponding fit
versus detuning $\epsilon$ for the upper wing, yielding $t=81\,
\mu$eV (black). (d) Same as (c), but for the lower wing, yielding
$t=64\,\mu$eV (red).}
\end{figure}

Next, we will analyze the linear differential conductance
$dI/dV_{sd}$ on the two high-conductive wings with respect to the
detuning $\epsilon$. With the help of Eq.~\ref{eq:coefficients}-\ref{eq:Gamma}
we are able to fit the function
$dI/dV_{sd}(\epsilon)$ and determine $t$ and also $\mid\alpha\mid^2$.
In Fig.~\ref{figure2} the differential conductance along
the top and the bottom wing is plotted versus the de\-tuning. As expected
$dI/dV$ is strongly peaked at zero de\-tuning and decays on the scale of $t$.
The solid curves are the corresponding fits, yielding $t=34 \,\mu$eV
and $t=40\,\mu$eV for the top and bottom wing, respectively. These
values are in excellent agreement with the tunnel coupling obtained
by analyzing the spacing of the two wings. The slightly asymmetric
shape of the conductance peak reflects an asymmetric coupling of the
double quantum dot to source and drain. The fits yield
$\Gamma_{d}/\Gamma_{s}=3.6$ and $6.7$ for top and bottom wing,
respectively. Knowing the tunnel coupling and using
Eq.~\ref{eq:coefficients}, we can now plot the probability of
finding the delocalized electron in, for example, the left dot,
$\mid\alpha\mid^2$, which is shown for convenience as an overlay
in  Fig.~\ref{figure2}(a).

To study the robustness of this approach, we will shortly focus on a
region with a larger tunnel coupling, for gate-voltages
around $V_{G1} \approx 2$~mV and $V_{G2}\approx 7$~mV (this is outside the
range shown in Fig.~\ref{figure1}(d)). An analysis of the
capacitances yields $C_1 \approx 78$~aF, $C_{2} \approx 64$~aF, and
$C_m \approx 19$~aF. In the analysis we have used the gate
capacitances obtained in Fig.~\ref{figure2}, $C_{G1}\approx 30$~aF,
and $C_{G2}\approx 25$~aF. Figure~\ref{figure3}(a) shows the two
high-conductance wings plotted versus $\epsilon$ and
$\Delta$. The spacing of the wings, $E_{\Delta}$, yields a tunnel
coupling of $t\approx 110\,\mu$eV and $U^{\prime}\approx
850\,\mu$eV, see Fig.~\ref{figure3}(b). The maximum differential
conductance along the two wings is plotted in Fig.~\ref{figure3}(c)
and Fig.~\ref{figure3}(d). A fit according to
Eq.~\ref{eq:coefficients}-\ref{eq:Gamma} yields $t=81\,\mu$eV
for the top wing and $t=64\,\mu$eV for the bottom wing. The
probability of finding the electron in the first dot, $\mid\alpha\mid^2$,
is again plotted as an overlay in Fig.~\ref{figure3}(a) for an
averaged value of $t=80\,\mu$eV. For the magnitude of $t$ the
agreement between the wing spacing and the conductance traces is not
quite as good as for the data presented in Fig.~\ref{figure2}.
However, taking into account the simplicity of the used approach,
the results are encouraging. In particular, we emphasize that the
wave\-function of the artificial molecule is extended over more than a
micron, neglecting the influence of impurities which are likely to
be present in the carbon nano\-tube.

%%%%%%%%%%%%%%%%%%%%%%%%%%%%%%%%%%%%%%%%%%%%%%%%%%%%%%%%%
%\section{Conclusions}
%%%%%%%%%%%%%%%%%%%%%%%%%%%%%%%%%%%%%%%%%%%%%%%%%%%%%%%%%

In this article, we have shown that electron delocalization in an
artificial molecule can directly be traced by electrical transport
measurements. In particular, this technique allows for mapping
molecular wave functions of the kind
$\ket{\Psi^{\pm}}=\alpha(\epsilon) \ket{\phi_1} \pm
\beta(\epsilon)\ket{\phi_2} $. Our results reflect the close
relationship of covalent bonding and efficient electronic
exchange, being one of the key challenges to be mastered in
molecular electronic devices \cite{Reed}. For the field of carbon
nano\-tube double quantum dots, the findings represent an additional
step towards the controlled engineering of quantum states. Future
experiments will have to address the dynamics of the electron spin
in carbon nano\-tube quantum dots, as these may be well-suited
building blocks for spin-based quantum computers \cite{Burkard}.

%%%%%%%%%%%%%%%%%%%%%%%%%%%%%%%%%%%%%%%%%%%%%%%%%%%%%%%%%%%%%%%%%%
\section{acknowledgments}
%%%%%%%%%%%%%%%%%%%%%%%%%%%%%%%%%%%%%%%%%%%%%%%%%%%%%%%%%%%%%%%%%%%
This work was obtained in collaboration with D. Loss and B. Coish.
Discussion with them are gratefully acknowledged. We also would like
to thank E. Bieri, A. Eichler, J. Furer, V.\ N. Golovach, L. Gr{\"u}ter,
G. Gunnarson and C. Hoffmann for assistance and discussions. We thank the
LMN of the Paul-Scherrer Institute for continuous support in
Si processing and we further acknowledge financial support from the Swiss
NFS, the NCCR on Nanoscale Science and the `C. und H. Dreyfus Stipendium'~(MRG).

%---------------------------------------------------------------------------
% Bibliography

\bibliographystyle{apsrev}

\end{document}